# All-Dielectric Photo-thermo-optical Metasurfaces for Thermal Landscaping at the Nanoscale

Gopal Narmada Naidu, Omer Can Karaman, Giulia Tagliabue

Abstract

Precise control of temperature fields at the micro- and nanoscale is essential for emerging applications in nanophotonics, catalysis, and microfluidics, yet remains difficult due to the diffusive nature of heat. While inverse-design algorithms have advanced thermoplasmonic metasurfaces, their extension to all-dielectric systems has not been explored. Here, an inverse thermal design framework is introduced for dielectric metasurfaces composed of thermo-optical amorphous silicon (a-Si) nanoresonators. By leveraging a precomputed library of absorption spectra as a function of geometry and temperature, target thermal profiles are directly mapped onto metasurfaces, enabling both uniform and complex temperature shaping. Unlike plasmonic platforms that require multi-resonator unit cells for tunability, dielectric nanoresonators provide intrinsic reconfigurability: at wavelengths where the thermo-optical coefficient is negligible (e.g., ~500 nm), absorption is temperature-invariant, whereas at other wavelengths it becomes strongly temperature-dependent, allowing illumination intensity to reshape the thermal landscape. This multifunctionality permits a single metasurface to yield distinct profiles under different excitation conditions without added structural complexity. As a proof of concept, photothermal catalysis on such metasurfaces is modeled, predicting over 30% enhancement in reaction rates. The presented framework establishes a scalable strategy for engineering nanoscale temperature fields with broad implications for catalysis, thermal management, and photothermal energy conversion.

Introduction

Precisely shaping temperature fields at the micro- and nanoscales is a key challenge in nanotechnology, with applications in photonics, nanochemistry, biology, and microfluidics.[1], [2], [3], [4], [5] Unlike electromagnetic fields, thermal fields are diffusion-driven, which makes spatial control difficult and pose challenges for applications requiring uniform or complex temperature distributions.[6] Thermonanophotonics, the study of photothermal effects in optical nanoantennas, has garnered significant attention in addressing this challenge. Most research to date has focused on thermoplasmonic systems, where metallic nanoparticles strongly absorb light and dissipate the energy as heat.[7] These studies have extensively examined both self- and collective-heating in such systems and, more recently have explored ultrafast or time-dependent photothermal effects for applications like photocatalysis.[8], [9], [10] However, recent advances in all-dielectric nanophotonics provide a compelling alternative for photothermal control.[11], [12] Resonant high-index dielectric and semiconductor nanostructures exhibit low-loss resonances below the bandgap, and strong absorption above the bandgap.[13], [14], [15], [16] Moreover, many of these materials have large thermo-optical coefficients (dn/dT), enabling efficient light–matter interactions along with substantial thermal tunability of their optical response. Furthermore, their low thermal capacity also leads to rapid photothermal responses.[17], [18] Notably, silicon nanoresonators have demonstrated exceptionally strong thermo-optical nonlinearities, in some cases surpassing plasmonic counterparts in terms of localized heating efficiency.[19], [20], [21]

While single dielectric nanoresonators have been studied in detail, extending thermo-optical control to large arrays remains challenging due to inter-resonator interactions and collective heating effects. In such systems, collective heating alters local absorption cross-sections and modifies individual resonator responses.[22] Recent work demonstrates that optimizing parameters such as resonator size, array periodicity, and illumination wavelength, yield non-trivial thermal profiles. For instance, spatially flat temperature distributions in heterogeneous arrays of silicon nanoresonators.[23] However, such brute-force parametric optimization is computationally expensive and impractical for designing metasurfaces with tailored temperature landscapes, motivating the need for inverse-design strategies.[24], [25] Beyond static optimization, thermo-optical resonators also offer opportunities for dynamic thermal reconfiguration.[21], [26], [27], [28] Because their optical response depends on both geometry and temperature, a single nanoresonator can, in principle, support multiple thermal output patterns. By changing the illumination wavelength or intensity, one nanoresonator could produce different heating effects, eliminating the need for complex multi-resonator unit cells to achieve tunability. Despite this potential, such tunable nanoscale thermal control remains largely unexplored.

In this work, we introduce the first inverse-design framework for shaping temperature fields in all-dielectric metasurfaces with strong thermo-optical response. Our approach combines a library of absorption spectra of amorphous silicon (a-Si) nanodisks - parameterized by geometry and temperature - with a matrix formulation of self- and collective heating to inversely map target temperature profiles to nanopattern designs. This framework enables a direct computational mapping of an arbitrary target temperature profile onto a spatially varying nanoresonator distribution. We demonstrate metasurfaces that produce either uniform or spatially complex thermal fields under uniform illumination thanks to the versatility of our static thermal shaping approach. Furthermore, by exploiting the wavelength-dependent thermo-optical coefficient of a-Si, we achieve temperature-invariant operation at a wavelength (~500 nm) where the TOC is negligible, and intensity-dependent thermal tuning at wavelengths where the TOC is appreciable. In other words, the same metasurface can exhibit no thermal change with varying intensity at one wavelength, yet show strong intensity-dependent heating at another wavelength – realizing multiplexed thermal functionality. Finally, as a proof of concept, we show that reshaping temperature profiles without altering absorbed power or average temperature can enhance photothermal catalytic reaction rates by over 30%, exploiting the nonlinear dependence of reaction kinetics on temperature. This mechanism, previously underutilized, offers a general route to boost catalytic performance and can extend to other supralinear photothermal processes such as water evaporation and separation.[29], [30], [31] Our approach thus unifies thermal precision, scalability, and dynamic reconfigurability, offering a versatile platform for nanoscale thermal management and functional photonic devices.

Theoretical model

We employ a numerical approach to solve the inverse problem of determining the heat source density—and consequently, the nanoresonator distribution—required to achieve a desired temperature field, T(x,y) across a metasurface. This approach has been previously developed and explored in literature for plasmonic systems [6], [32], and is here adapted for designing all-dielectric metasurfaces. The theoretical framework underpinning this approach is based on the temperature increase $\Delta T_i$ of the i-th nanoresonator, which consists of two components: the self-heating term ($\Delta T_i^s$) and the collective heating term ($\Delta T_i^{ext}$).[33]

$$\Delta T_i = \Delta T_i^s + \Delta T_i^{ext}$$

The self-heating contribution stems from the nanoresonator's own absorption of light - the temperature contribution of the i-th nanoresonator itself to the temperature rise is given by

$$\Delta T_i^s = \frac{\sigma_{abs} I}{4\pi R_{eq} \beta \kappa} = \frac{(absorptance) P}{4\pi R_{eq} \beta \kappa} = \frac{Q_i}{4\pi R_{eq} \beta \kappa}$$

where $\sigma_{abs}$ is the absorption cross-section of the nanoresonator, I the intensity of the incident light, P the incident light power and $R_{eq}$ the equivalent radius (the size equal to the radius of a sphere with the same volume as the nanoresonator under study). The dimensionless heat capacity coefficient β accounts for arbitrary resonator geometries and, for nanodisks, is given by: $\beta = \exp\{0.040 - 0.0124 \ln\left(\frac{D}{t}\right) + 0.0677 \ln^2\left(\frac{D}{d}\right) - 0.00457 \ln^3\left(\frac{D}{d}\right)\}$ .[34] Here D and t are the nanodisk diameter and thickness respectively. The thermal conductivity of the surrounding medium is denoted by κ is the thermal conductivity of the media in which the particles are embedded. When considering an interface which separates two isotropic media with thermal conductivities $\kappa_1$ and $\kappa_2$, an average thermal conductivity $\left(\bar{\kappa} = \frac{\kappa_1 + \kappa_2}{2}\right)$ of the two environmental media is defined. [35]

The collective heating term - the temperature contribution due to the remaining N-1 resonators around the i-th nanoresonator - is defined as

$$\Delta T_i^{ext} = \sum_{\substack{j=1 \\ j \neq i}}^{N} \frac{Q_j}{4\pi \bar{\kappa} r_{ij}}$$

Hence, the total temperature increase becomes: $\Delta T_i = \Delta T_i^s + \Delta T_i^{ext} = \frac{1}{4\pi R_{eq} \beta \bar{\kappa}} Q_i + \sum_{\substack{j=1 \\ j \neq i}}^{N} \frac{1}{4\pi \bar{\kappa} r_{ij}} Q_j$

This expression can be rewritten in matrix form as **T** = A **Q**

where A is an NxN matrix with $A_{ij} = \begin{cases} \frac{1}{4\pi R_{eq} \beta \bar{\kappa}}, & i = j \\ \frac{1}{4\pi \bar{\kappa} r_{ij}}, & i \neq j \end{cases}$

For a given target temperature distribution **T**, the corresponding thermal power distribution can be determined by inverting matrix A and obtaining **Q** = A$^{-1}$**T**. [6], [32] While the inversion algorithm always produces a heat source distribution for any temperature map, we note that certain input temperature distributions can lead to non-physical solutions. Specifically, some computed heat source distributions may contain negative values, indicating the presence of heat sinks rather than sources. Another prohibitive case occurs when the target temperature distribution includes zero values within the simulation domain. This arises because, once any nonzero heat source is introduced, heat spreads throughout the system, making it impossible to maintain strictly zero temperature at any point. These constraints stem from the mathematical properties of A and its invertibility. Therefore, when designing target temperature distributions, careful physical considerations must be applied to ensure feasible and realistic solutions.

While designing the photothermal metasurfaces, it is also important to note that for a point source, the intensity of electromagnetic wave decays by $1/r^2$, while the resulting temperature field decays more slowly as $1/r$. Thus, in the absence of optical coupling, the optical response of the array can be approximated by that of the individual resonators, significantly reducing computational demands. Unlike optical fields, the temperature field generated by a large array of nanoheaters cannot be simplified in the same manner. When designing the temperature field, the thermal contributions from neighboring particles must be carefully considered. However, when the interparticle distance is less than four to five times the nanoresonator diameter, electromagnetic coupling becomes significant and must be accounted for. We address this by using the unit cell approach, which captures near-field electromagnetic interactions between nanoresonators. This unit cell approach enables a direct mapping of thermal power density to the absorptance of specific optical nanoresonators placed in an array.

We subsequently consider a metasurface consisting of a-Si nanodisks with a height of 100 nm arranged in a square lattice with pitch set to 350 nm, to avoid any diffractive mode in the visible to near-infrared regime. The nanodisks' radius R is varied from 50 nm to 150 nm. Using the unit cell approach and incorporating the temperature-dependent refractive index of Si, we calculated (COMSOL®) the absorption spectra of Si nanodisks of varying radii and temperature. The refractive index of Si was obtained from temperature-dependent ellipsometric measurements and validated against experiments for nanoresonator design.[17] The dielectric functions and other material parameters used in our simulations are listed in **Table S1**. By transforming the desired thermal power density distribution **Q(x,y)**, obtained from our inversion algorithm for a given temperature profile **T(x,y)**, into an absorption cross-section distribution $\sigma_{abs}(x_i, y_i)$ and referencing the spectral library generated through the unit cell approach in COMSOL, we constructed a detailed nanoresonator metasurface, R(x,y). This represents, to the best of our knowledge, the first application of such an inversion algorithm for dielectric metasurfaces with a significant thermo-optical effect. Thermo-optical effects can dynamically alter the absorption spectrum of the nanoresonators and, consequently, their photothermal conversion, creating a closed feedback loop. As a result, $\sigma_{abs}$ becomes a function of both the nanodisk radius R and the temperature T. Therefore, the designed temperature profiles can now be dynamically tuned via changing either the irradiation intensity or wavelength.

Results and discussion

1. Photothermal metasurface design and algorithm validation

A schematic diagram illustrating the conversion between thermal power density distribution and temperature distribution in a metasurface composed of identical nanoresonators is shown in **Fig. 1A.** When a uniform thermal power density is applied, thermal diffusion leads to a non-uniform temperature distribution (**Fig. 1A**, top panels), with collective heating producing a peak temperature at the center of the array. To achieve a uniform temperature distribution across the metasurface (**Fig. 1A,** bottom panels), the thermal power density must be weaker in the center and stronger at the edges to counteract diffusion effects, as also shown in previous studies.[6], [32] Beyond uniform heating, more complex temperature distributions such as linear temperature gradients and parabolic profiles, can also be realized via power density engineering. Shaping the photoexcitation beam to spatially modulate the light intensity on a uniformly distributed nanoresonator array is the most intuitive approach to realize arbitrary temperature landscapes. However, its implementation requires a spatial light modulator and, due to the diffraction limit, offers relatively low spatial resolution for power density control.[36], [37] An alternative and more effective strategy is to adjust local dissipation under uniform illumination, either by varying the nanoresonator density or by tuning their absorptance properties.

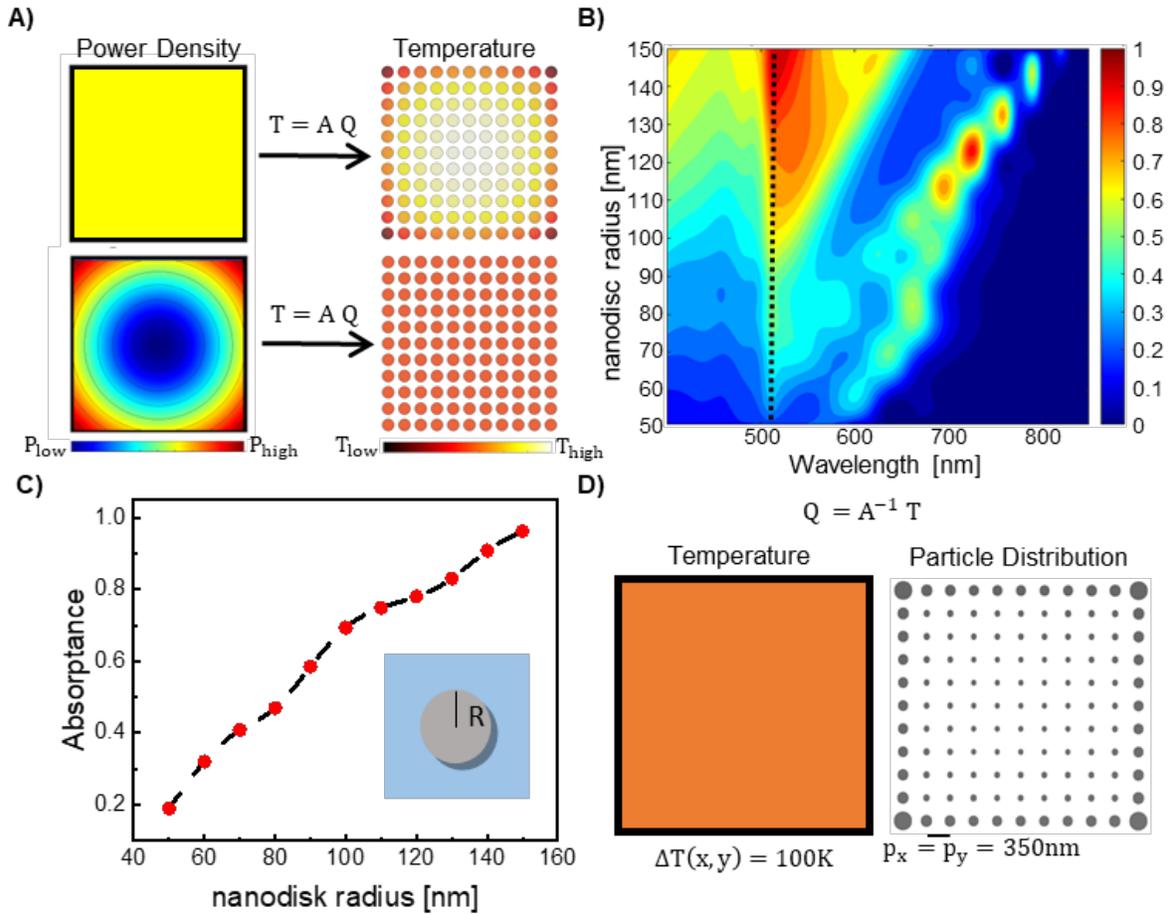

**Figure 1.** A) Uniform and inverse-gaussian thermal power density distributions and their corresponding temperature distributions obtained over a 4 μm x 4 μm a-Si metasurface composed of a-Si nanodisks (R = 150 nm, H = 100 nm, P = 350 nm) on a silica substrate. B) Calculated absorptance spectra of a-Si nanodisks (H = 100 nm) at room temperature (T = 293 K) with varying radii (R = 50-150 nm), obtained using a unit cell approach (P = 350 nm). C) Absorptance of a-Si nanodisks at a target wavelength of 520 nm and room temperature (T = 293 K), as a function of radius. **Inset:** Schematic illustration of the simulated a-Si nanodisk unit cell. D) Particle distribution for an 11x11 array of a-Si nanodisks on silica (4 μm x 4 μm metasurface) to achieve a uniform temperature increase of ΔT(x,y) = 100 K.

**Fig. 1B**, presents the calculated absorptance spectra of a-Si nanodisks at room temperature (293K) on a silica substrate, with height, H of 100 nm and varying radii, R using the unit cell approximation (pitch, P = 350 nm). As expected, the resonance behavior of the nanodisks strongly depends on their size. Two distinct trends emerge: at 520 nm, all disks exhibit an absorption peak,

whose magnitude increases with disk radius, as shown more clearly in **Fig. 1C**; meanwhile a second resonance mode, red shifts strongly with increasing nanodisk radius, which we will discuss later.

To make a full computational validation of the inverse-design algorithm feasible using COMSOL, we initially focus our attention on a small-footprint metasurface, composed of 11x11 a-Si nanodisks corresponding to ~ 4 µm x 4 µm metasurface. As shown in **Fig. 1D**(left), we target a uniform temperature increase across the metasurface, $\Delta T(x,y) = 100K$. The optimized a-Si nanoresonator distribution obtained from the inverse algorithm under uniform illumination at 520 nm and 0.25 mW/µm$^2$ is presented in **Fig. 1D**(right). The nanoresonator distribution consists of small nanodisks across the center of the array, larger nanodisks along the edges and the largest nanodisks at the corners (exact nanoresonator size distribution, R(x,y) in **Fig. S1**). We simulated the temperature distribution of this photothermal metasurface in COMSOL using coupled optical and thermal equations obtaining excellent agreement with the inverse-design algorithm prediction (**Note S1**).

2. Temperature-invariant operation of the metasurfaces

The verification metasurface design leveraged solely the shape-dependent optical response of the nanoresonators, neglecting variations in refractive index with temperature (dn/dT, or thermo-optic effect). However, thermo-optic (TO) effects can dynamically shift the absorption spectrum of the nanoresonators, altering their photothermal conversion and creating a feedback loop that continuously modifies the pre-designed temperature distribution.[17], [22], [23] Therefore, the target temperature profile, designed by controlling photothermal conversion and heat diffusion in the metasurface, remains valid only under specific irradiation intensity and wavelength conditions. We now extend our analysis to include the temperature-dependent response of these dielectric nanoresonators.

We calculated the absorptance spectra of the a-Si nanodisks at different temperatures, up to 600$^0$C, using experimentally measured complex refractive index values.[17] As shown in **Fig. 2A** for a-Si nanodisks with radius of 150 nm, the absorptance spectrum exhibits major variations with temperature across the 600–900 nm wavelength range, while remaining temperature-invariant around ~ 500 nm. This is indeed due to the wavelength-dependent TO coefficient of a-Si, which is near-zero at 488 nm.[17] Being due to an intrinsic material property, this behavior is consistent across nanoresonators of varying radii (**Fig. 2B**). As a result, under uniform excitation at 520nm, the previously designed metasurface maintains an intensity-invariant, uniform temperature distribution (**Fig. 2C**), with only the absolute value of the temperature scaling with irradiation intensity (0.25, 0.5, 1, and 1.5 mW/µm$^2$). This result highlights the importance of characterizing the wavelength-dependent TO coefficient of materials, towards identifying possible regimes of TO-free operation for temperature-invariant metasurfaces.

To further illustrate the potential of the inverse-design algorithm for nanoscale temperature landscaping, beyond the diffraction limit, we targeted the creation of multiple hotspots while counteracting the effect of collective heating on a larger metasurface (10 µm x 10 µm metasurface). In **Fig. 2D**, we show the metasurface designed to achieve a sinusoidal temperature profile in both x- and y- directions (exact nanoresonator size distribution in **Fig. S2**). Specifically, at an irradiation intensity of 3 mW/µm² at 520 nm, we achieved alternating hot ($\Delta T = 110\ ^0C$) and cold ($\Delta T = 80\ ^0C$) spots on the metasurface, highlighting the capability to precisely control temperature at the sub-micron scale. This metasurface still exhibits intensity- and temperature-invariant behavior in terms of the normalized spatial temperature distribution. However, the absolute temperature gradient scales with irradiation intensity.

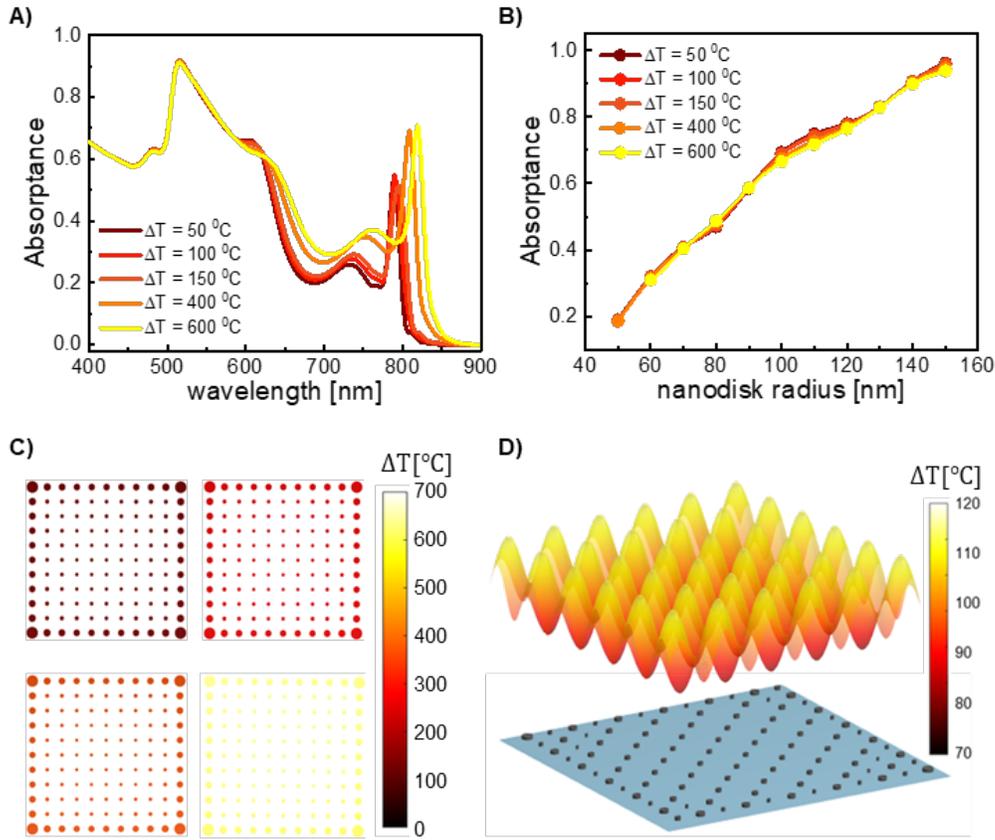

**Figure 2.** A) The calculated absorptance spectra of a-Si nanodisk array (R = 150 nm; H = 100 nm; P = 350 nm) at different temperatures (70, 120, 170, 400 and 600°C) B) The absorptance at 520 nm excitation of nanodisks of varying radii as a function of temperature. C) Temperature invariant performance (at 100, 200, 400 and 600°C) of the designed metasurface at increasing intensities of irradiation of 0.25, 0.5, 1, and 1.5 mW/µm² respectively. D) Designed 10 µm x 10 µm a-Si nanodisk on silica metasurface for achieving a sinusoidal temperature profile in both the x- and y- directions at 3 mW/µm² uniform illumination with a 520 nm laser (exact particle distribution in SI, Fig. S2).

3. Intensity-dependent multiplexing of spatial temperature distributions on silicon metasurfaces

We now explore the opportunities offered by strong TO effects in the 600-900 nm spectral range (**Fig. 2B**) for temperature landscaping.

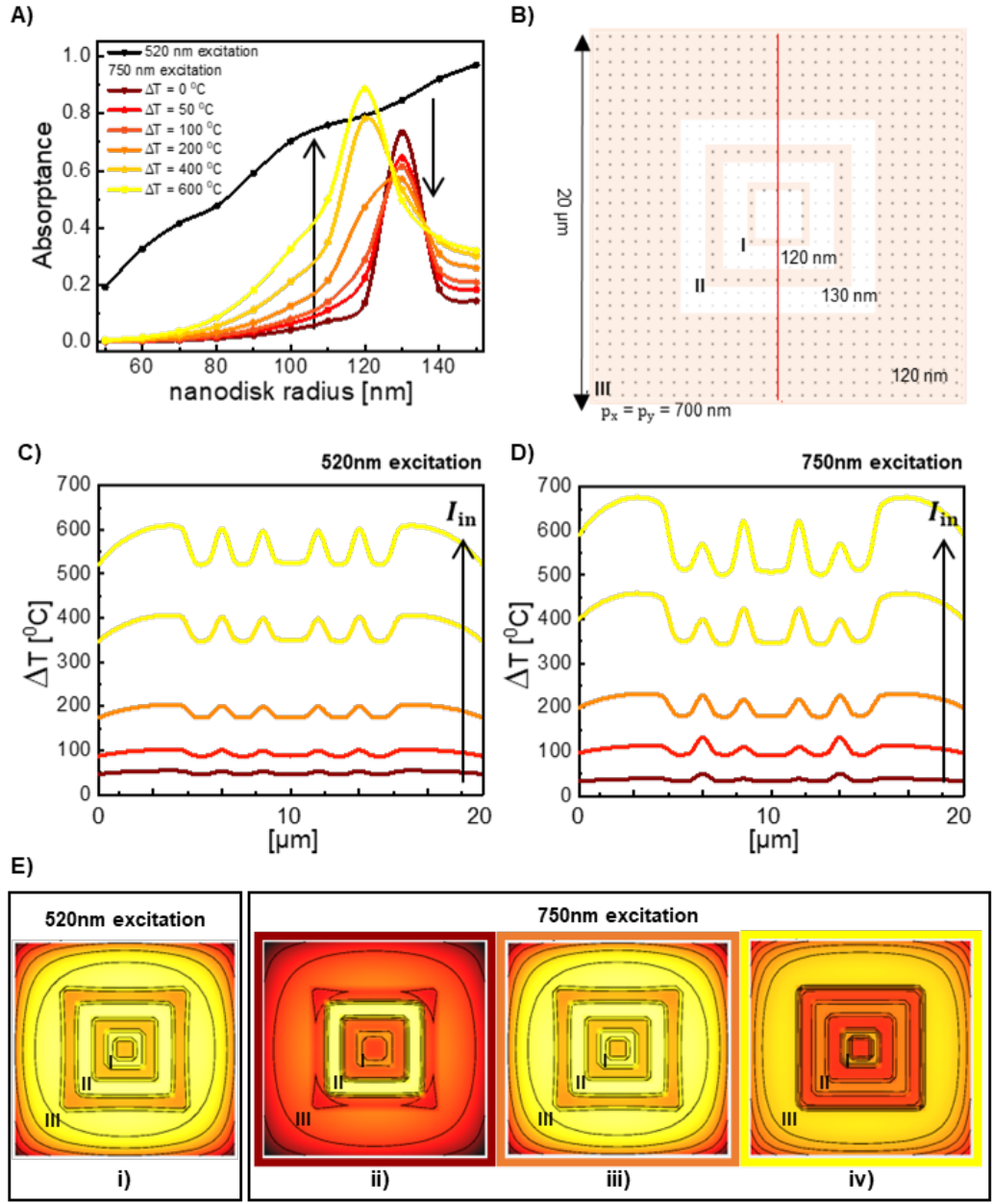

**Figure 3.** A) The absorptance of the a-Si nanodisks of varying radii at 520 nm and 750 nm excitation as a function of temperature increase ($\Delta T$). B) Schematic of a 20 μm x 20 μm a-Si nanodisk metasurface on silica, designed to achieve a radially oscillating temperature profile. Temperature profile across a radial cross-section of the metasurface under increasing irradiation intensity at C) 520 nm and D) 750 nm excitation. E) Normalized spatial temperature distribution over the entire metasurface at i) 520 nm and ii-iv) 750 nm excitation, demonstrating the multiplexing of spatial temperature distributions via the thermo-optical effect of a-Si.

**Fig. 3A** shows the absorptance of a-Si nanodisks with varying radii at excitation wavelengths of 520 nm (same as **Fig. 2C**) and 750 nm as a function of their temperature increase. As discussed earlier, at 520 nm excitation, the absorptance remains invariant with respect to intensity and hence temperature. However, at 750 nm excitation, the nanodisks display a thermally tunable response:

at room temperature, 130nm nanodisk absorbing strongly due to its narrow resonance at 750 nm. As the temperature increases, this resonance undergoes a redshift, causing the 130 nm nanodisk to eventually be off resonance from $\Delta T > 200\ °C$. Simultaneously, the 120 nm nanodisk has a resonance initially at 720 nm. This also red shifts with temperature, eventually making it on-resonance with 750 nm excitation for $\Delta T > 200\ °C$. Consequently, as temperature increases, the absorptance of the 130 nm nanodisks decreases, while that of the 120 nm nanodisks increases, demonstrating a dynamic, thermally tunable optical response that can be used for reconfigurable temperature landscaping.

**Fig. 3B** illustrates a 20 μm × 20 μm a-Si nanodisk metasurface on a silica substrate composed of nanodisks with radii of 50 nm, 120 nm, and 130 nm. This structure is engineered to achieve a spatially oscillating radial temperature profile while counteracting the effects of collective heating, thereby maintaining a relatively flat temperature distribution. **Fig. 3C-D** show the temperature distribution along a radial cross-section (x- or y- cross-sections, red line in **Fig. 3B**) of the metasurface for increasing irradiation intensities at 520 nm and 750 nm excitation, respectively. At 520 nm excitation, as discussed previously, increasing the irradiation intensity enhances the temperature gradients across the metasurface while preserving the spatially oscillating radial temperature profile. This is very evident in **Fig. 3E.i**, which shows the normalized temperature profile across the metasurface remaining unchanged regardless of irradiance intensity as a result of the near-zero thermo-optical coefficients of a-Si in this wavelength regime.

Instead, at 750 nm excitation the thermo-optical coefficient of a-Si is significantly higher, making the spatial temperature distribution strongly intensity-dependent. At lower intensities ($\Delta T_{avg} \approx 50 - 100\ °C$, **Fig. 3D, dark red and red curves**), heating is predominantly localized in region II, composed of 130 nm radius nanodisks. As the irradiation intensity increases ($\Delta T_{avg} \approx 200\ °C$) heating becomes more uniform across all nanodisks, creating a temperature profile similar to that observed under 520 nm excitation (**Fig. 3D, orange curve**). Upon further increasing the intensity, a distinct pattern emerges: Regions I and III experience selective heating, while Region II remains at a relatively lower temperature (**Fig. 3D, yellow curves**). This intensity-dependent multiplexing of the spatial temperature distribution under 750 nm excitation is clearly observed in **Fig. 3E. ii-iv**, which displays the normalized temperature distributions. Such spatially controlled heating as a function of light wavelength and/or intensity has potential applications in precise fluid manipulation in microfluidic devices[5] and influencing reaction kinetics.[1] To demonstrate this potential, the enhancement of reaction kinetics through temperature shaping will be discussed further in the next section.

4. Enhancing the reaction kinetics by temperature shaping

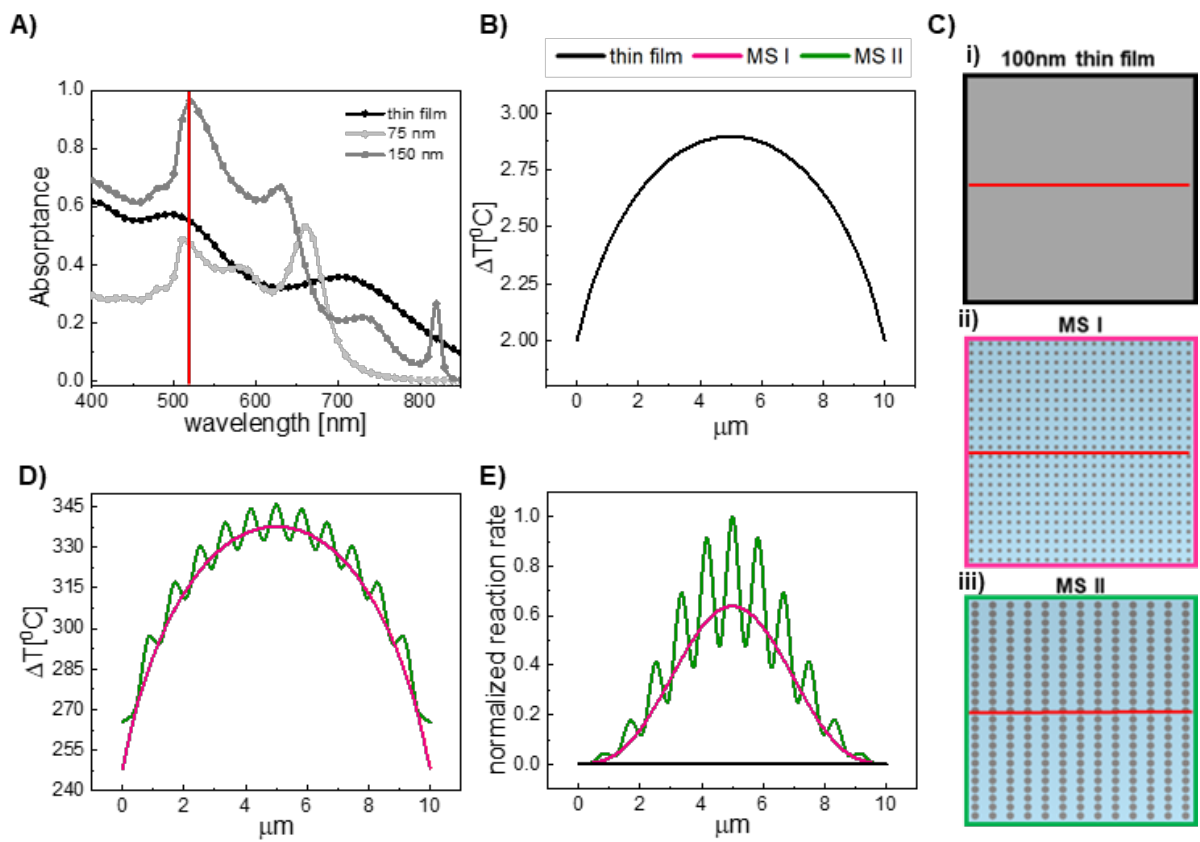

**Figure 4.** A) Calculated absorptance spectra of a 100 nm-thick a-Si film and a-Si metasurfaces composed of nanodisks with radii of 75 nm and 150 nm at room temperature. B) Simulated temperature distribution in a 100 nm-thick a-Si film (10μm x 10μm) illuminated with a 0.3 mW/μm² laser intensity at 520nm. C) Schematic of the 10 nm a-Si thin film and the two metasurfaces: MS I – a uniform array of nanodisks (radius: 75 nm, periodicity: 350 nm) and MS II – an optimized metasurface composed of 150 nm nanodisks designed to maintain the same absorbed power and average temperature as MS I at a given intensity. D) Corresponding temperature profiles for MS I and MS II. E) Normalized reaction rates for a photo-assisted thermocatalytic reaction on the thin film, MS I and MS II.

For photo-assisted thermocatalytic reactions—such as gas-phase photomethylation of $CO_2$ in hydrogen—the reaction rate is strongly correlated with substrate heating under illumination (photothermal heating). In particular, methanation rates have been shown to increase exponentially with temperature.[38]

$$k = Ae^{-\frac{E_a}{RT}}$$

where, k is the rate constant, A is the pr-exponential factor, $E_a$ is the activation energy of the reaction, and R is the universal gas constant.

As shown in **Fig 4A**, a 100 nm-thick a-Si film exhibits an absorptance of ~ 0.5 at 520 nm. Under continuous-wave (CW) illumination with a 520 nm laser at an intensity of 0.3 mW/μm², a 10μm x 10 μm film develops a gaussian temperature profile with a maximum temperature increase ($\Delta T_{max}$) of only ~3 °C (**Fig 4B**), limiting the achievable thermally driven reaction rate. However, by nanopatterning the film into a uniform metasurface of nanodisks (radius: 75 nm, periodicity: 350 nm, **Fig. 4C, ii**), we achieve a dramatic increase in $\Delta T_{max}$ to ~300 °C - even though the absorbed power at 520 nm excitation remains unchanged (**Fig 4D, pink curve**). This enhancement results from a reduction in the system's effective thermal conductivity due to the metasurface patterning, which disrupts heat dissipation pathways and localizes thermal energy more effectively. This is evident from the equation, $\kappa_{eff} = f \cdot \kappa_{a-Si} + (1 - f)\kappa_{air}$, where f is the volume fraction of a-Si to that of air.

To highlight the role of temperature shaping on the reaction kinetics, a second metasurface (MS II) was designed using larger nanodisks (150 nm radius, **Fig. 4C, iii**) optimized to maintain the same absorbed power and average temperature as MS I for a given intensity. However, unlike the temperature distribution in MS I, MS II produces a highly modulated thermal landscape, featuring nanoscale "hot spots" of significantly elevated local temperatures (**Fig 4D, green curve**). These spatial temperature variations have a profound impact on catalytic activity. Assuming the same pre-exponential factor, $A = 1.98 \times 10^6$ mmol/g.h for both MS I and MS II, we computed the spatially integrated reaction rate for each design. As shown in **Fig. 4E**, MS II exhibits a normalized reaction rate more than 30% higher than that of MS I, despite having the same average temperature. This performance boost stems from the exponential sensitivity of reaction rates to temperature, enabling localized regions of higher temperature to disproportionately enhance the overall reaction rate. We used an activation energy $E_a = 54.5$ kJ/mol consistent with prior reports on $CO_2$ photomethanation.[38] Importantly, this reveals that not only the average temperature but also its spatial distribution plays a critical role in determining photothermal catalytic efficiency. Harnessing this inherent but previously overlooked nonlinearity offers a powerful strategy for designing high-throughput catalytic surfaces. Moreover, this mechanism can enhance a broad range of photothermal processes with supralinear temperature dependence, such as water evaporation [29] and separation techniques [31].

Conclusions

In this work, we have developed and validated an inverse-design algorithm for all-dielectric photothermal metasurfaces capable of sculpting arbitrary temperature landscapes with high spatial resolution. By solving the inverse problem of mapping desired temperature profiles to thermal power distributions and then linking these to nanoresonator geometries through a unit cell approach, we demonstrated full control over static and dynamically reconfigurable temperature fields at the sub-micron scale. Validation with a-Si nanodisk metasurfaces capable of achieving uniform and complex temperature distributions under uniform illumination demonstrated excellent agreement between theoretical predictions and COMSOL simulations, confirming the accuracy of our model.

By carefully selecting the excitation wavelength (520 nm) corresponding to the near-zero thermo-optical (TO) coefficient of a-Si, we demonstrate metasurfaces whose temperature profiles remain invariant under varying light intensities, highlighting the possibility of designing robust, predictable temperature fields. While by leveraging the strong TO response at 750 nm, we achieved dynamic, intensity-dependent reshaping of temperature profiles, enabling spatial multiplexing of heating patterns. This tunability opens new opportunities for active thermal control, beyond static temperature fields. Additionally, we also demonstrate that engineered temperature shaping can significantly enhance local heating efficiency at constant absorbed power, leading to large increases in thermally activated reaction rates. This highlights the practical relevance of our approach for catalysis and other heat-driven processes. Overall, this study introduces a versatile and scalable strategy for nanoscale thermal management using dielectric metasurfaces. Beyond fundamental interest, our findings offer promising routes for applications in microfluidic control, thermocatalysis, optothermal actuation, and reconfigurable photothermal devices.

Research Funding Statement

G.N.N. and G.T. acknowledge the support of the Swiss National Science Foundation (Starting Grant No. 211695).

Author Contribution Statement

G.T supervised all aspects of the project. G.N.N. built the algorithm and the transient simulation model and performed all the simulations. O.C.K. played an active role in the discussions and conceptualization. G.T. and G.N.N. wrote the paper, with input from all the other authors.

Conflict of Interest Statement

The Authors declare no competing interests.